\documentclass[english,aps,prb,superscriptaddress,reprint,twocolumn,longbibliography]{revtex4-2}

\usepackage[utf8]{inputenc}
\usepackage[T1]{fontenc}
\usepackage{amsmath}
\usepackage{amssymb}
\usepackage{amsfonts}
\usepackage{bbold}
\usepackage{tikz}
\usepackage{xfp}
\usepackage{float}

\usepackage[labelfont=bf, font=small, hypcap=false]{caption}

\usepackage{color}

\usepackage{graphicx}
\usepackage{dcolumn}
\usepackage{bm}
\usepackage{commath}
\usepackage{upgreek}
\usepackage{siunitx}

\AtBeginDocument{\RenewCommandCopy\qty\SI}
\usepackage{natbib}
\usepackage[english]{babel}

\usepackage[bottom=2cm,top=2cm, left=1.5cm, right=1.5cm]{geometry}
\usepackage{booktabs}
\colorlet{linkequation}{blue}
\usepackage[normalem]{ulem}
\usepackage{xcolor}
\usepackage{hyperref}

\hypersetup{
    colorlinks=true,
    linkcolor=blue,
    citecolor=blue,
    urlcolor=blue
}

\begin{document}
\preprint{APS/123-QED}

\title{Field-derivative torque induced magnetization reversal in ferrimagnetic $\rm \bf{ Gd_{\frac{3}{2}}}\bf{Yb_{\frac{1}{2}}BiFe_{5}O_{12}}$}

\author{Pratyay Mukherjee}%
\email{23dr0111@iitism.ac.in}
\affiliation{Department of Physics, Indian Institute of Technology (ISM) Dhanbad, Dhanbad, Jharkhand, India, 826004}

\author{Arpita Dutta}
\affiliation{School of Physical Sciences, National Institute of Science Education and Research, An OCC of HBNI, Jatni, 752 050 Odisha, India}

\author{Somasree Bhattacharjee}
\affiliation{Department of Physics, Indian Institute of Technology (ISM) Dhanbad, Dhanbad, Jharkhand, India, 826004}

\author{Shovon Pal}
\affiliation{School of Physical Sciences, National Institute of Science Education and Research, An OCC of HBNI, Jatni, 752 050 Odisha, India}

\author{Ritwik Mondal}
\email{ritwik@iitism.ac.in}
\affiliation{Department of Physics, Indian Institute of Technology (ISM) Dhanbad, Dhanbad, Jharkhand, India, 826004}

\date{\today}

\begin{abstract}  
 Understanding the mechanism of spin switching in ferrimagnets via the excitation of THz pulses holds promise for future-generation magnetic memory devices. Such spin switching can be accomplished by the Zeeman torque exerted by the THz pulses on the magnetic spins. Theoretical and experimental works have established that the field derivative of a terahertz pulse also exerts a torque, field derivative torque (FDT). Here, we investigate the role of the FDT in the spin switching in ferrimagnetic $\rm Gd_{\frac{3}{2}}Yb_{\frac{1}{2}}BiFe_{5}O_{12}$ using a computational approach. Our results foresee that the spin switching in the presence of the FDT requires less THz magnetic fields than the spin switching without the FDT. Without the FDT terms, the spin switching in the considered system requires an extremely high magnetic field. Furthermore, we compute the switching and non-switching contour diagrams to show that the FDT tremendously enhances the possibility of spin switching. These results not only shed light on the significance of the FDT in magnetization switching but also suggest materials where the switching effect is pronounced.
\end{abstract}

\maketitle

\section{Introduction}
The discovery that magnetization can be manipulated using ultrafast pulses has paved the way for the field of ultrafast terahertz spintronics, with the ultimate goal of enabling faster and energy-efficient magnetic memory devices~\cite{Beaurepaire1996,Back1998}. At the heart of ultrafast magnetism and THz spintronics lies the quest for terahertz spin switching—reversing the orientation of spins in magnetic materials on timescales of femtoseconds (fs) and picoseconds (ps). Unlike conventional magnetic switching, which relies on slower precessional dynamics driven by magnetic fields or spin-transfer torques, THz spin switching promises a fundamentally faster and more compact route to information processing~\cite{Stanciu_Hansteen2007,Kirilyuk2010,Kampfrath2013,Nemec2018,Kampfrath2011,John2017,Li2022APL}. 
It has been demonstrated experimentally that fs optical pulses can induce all-optical switching in ferrimagnetic materials~\cite{Yang17,Radu2011,Kimel2007-Review,Mishra2021,yamamoto2015ultrafast,Kirilyuk2010,Koopmans2010,Beens2019}. In particular, experiments using circularly polarized ultrafast laser pulses have shown that non-thermal magnetization reversal is achievable in ferrimagnets~\cite{Stanciu2007,Kimel2005,Shi2023}. Unlike antiferromagnetic materials, ferrimagnets exhibit a non-zero net magnetization due to antiferromagnetically coupled magnetic sublattices, a property that enables magnetization switching on picosecond timescales~\cite{Sonke2012,Zhang2025,Vovk2025,Blank2024,Schlauderer2019}.

Such experimental results have been explained via the multiscale modeling using the atomistic spin dynamics together with the two-temperature model~\cite{Radu2011,ostler12,Kazantseva2008,Kazantseva2007,vahaplar12}. The dynamics are typically governed by the stochastic Landau-Lifshitz-Gilbert (LLG) equation, incorporating thermal fluctuations and exchange interactions at the atomic scale~\cite{Atxitia2009,Chimata2015,Evans2014,Evans2012,florian2022}. The LLG spin dynamics describes the temporal evolution of magnetization through two primary torque contributions: A precessional torque, which causes the magnetization vector to precess around the effective magnetic field, and a damping torque, which leads to energy dissipation and eventual alignment of the magnetization with the field. The strength of the damping torque is characterized by the Gilbert damping parameter, a dimensionless constant that quantifies how rapidly the system loses energy to its surroundings~\cite{landau35,landau1982mechanics,gilbert04}. From a theoretical standpoint, attempts to derive the LLG equation starting from the Dirac–Kohn–Sham (DKS) framework~\cite{Rajagopal1978,macdonald79,eschrig99,kraft95,crepieux01,Mondal2015a,ado2024}—in order to incorporate relativistic spin–orbit coupling effects have led to the emergence of an additional torque term, referred to as the field-derivative torque (FDT)~\cite{Mondal2016,Mondal2018PRB}. Note that the LLG equation has been extended to incorporate other spin torque effects as well e.g., spin transfer torque~\cite{Ralph2008,slonczewski96,Berger1996}, spin-orbit torque~\cite{Gambardella2011,Manchon2019RMP}, optical spin-orbit torque~\cite{tesarova13,Huang2024,mondal2021terahertz}, and inertial spin torque~\cite{Ciornei2011,neeraj2019experimental,unikandanunni2021inertial,Mondal2017Nutation,Mondal2020nutation,De2025experiment}.

It has been predicted that the FDT terms cannot be neglected for a system with higher Gilbert damping and excited by a THz pulse~\cite{Mondal2019PRB,Blank2021THz}. Numerical studies on antiferromagnetic systems suggest that FDT enhances the amplitude of the magnon oscillations~\cite{Mondal2019PRB}. Additionally, it introduces an extra phase factor in the magnon oscillation response e.g., the phase with FDT terms leads by $\pi/2$ in antiferromagnetic CoO and NiO~\cite{Mondal2019PRB}. The experimental realization of such FDT terms have been reported recently in rare-earth, Bi-doped iron garnet $\rm Gd_{\frac{3}{2}}Yb_{\frac{1}{2}}BiFe_{5}O_{12}$, a ferrimagnetic material with Gilbert damping 0.02 and the Kaplan-Kittel exchange resonance frequency 0.48\,THz~\cite{Dutta2025,Dutta2024}. Using ultrafast nonlinear magnetization dynamics, it was found that the observed nonlinear magnetic response cannot be fully attributed to magnetization precession driven solely by the Zeeman torque (ZT). Including a FDT alongside the ZT accurately reproduces the experimental results~\cite{Dutta2025}. Despite the recent progress in FDT, the experimental and theoretical literature is rather limited. Further investigations are essential to fully understand its contribution to ultrafast spin dynamics, switching, and to harness its potential in high-speed magnetic switching applications.    

In this study, we investigate the influence of the FDT on ultrafast magnetization switching dynamics in ferrimagnetic $\rm Gd_{\frac{3}{2}}Yb_{\frac{1}{2}}BiFe_{5}O_{12}$, triggered by excitation with a THz pulse. Using a circularly polarised THz pulse, we simulate the LLG spin dynamics without and with FDT terms for antiferromagnetically coupled spins. Our results show that the inclusion of FDT significantly lowers the threshold of the THz magnetic field needed for magnetization switching compared to when only the ZT is considered. For example, magnetization switching is not observed with a THz pulse of temporal width $\sigma$ = 1 ps and a peak magnetic field of 6.35 T when only the ZT is considered; however, the inclusion of the FDT terms enables successful magnetization reversal under the same conditions. We further observe that, in the absence of the FDT, a significantly higher magnetic field of 24.35 T is required to achieve magnetization switching. The switching diagrams, consisting of the relationship between the magnetic field, pulse temporal width, and the resulting magnetization switching behavior, have been computed. These switching diagrams indicate the existence of distinct switched and non-switched regions, both with and without the inclusion of FDT terms. Notably, the switched regions expand when FDT effects are considered; however, the diagrams still exhibit a periodic pattern of alternating switched and non-switched areas. We note that the magnetization switching induced by FDT is not significantly enhanced when driven by a linearly polarized THz pulse. 

We argue that the torques exerted on the magnetization by the circularly polarized THz pulse and the FDT interact more effectively for switching as compared to those generated by a linearly polarized THz pulse. This enhanced interplay arises due to the continuously rotating magnetic field vector of the circularly polarized pulse, which applies a torque that remains transverse to the magnetization throughout the pulse duration. Such a torque direction promotes coherent precessional motion, aiding the switching process. In contrast, a linearly polarized pulse applies a torque that oscillates back and forth along a fixed axis, resulting in partial cancellation over time and less efficient energy transfer to sustain the spin precession. The dynamic nature of the circular polarization better complements the FDT, facilitating more robust and deterministic magnetization switching.

The article is organized as follows: Section II outlines the computational methods employed discussing the details of LLG simulations with the FDT, modeling the THz pulse and material specification; Section III describes the results and discussions related to magnetization switching without and with FDT. Section IV concludes the results with a discussion on the potential implications for future technological advancements.

\section{Computational Details}
\subsection{Simulation of LLG spin dynamics}
To investigate magnetization reversal, we have simulated the magnetization dynamics using Landau–Lifshitz–Gilbert (LLG) equation both in the presence and absence of the FDT term. Although the LLG equation is fundamentally semiclassical, it remains remarkably effective in predicting spin dynamics on ultrafast timescales, even down to the picosecond regime~\cite{lambert14,ostler12,MorenoPRB2017,Moreno2019PRB,Markus2024}. The equation takes the following form:
\begin{align}
\label{Eq1}
\dot{{\bf m}}_{i} = -\frac{\gamma_i}{1+\alpha_i^2} \,{\bf m}_{i} \times \left[{\bf B}_{i}^{\rm eff} + \frac{{\alpha_i}}{\vert {\bf m}_{i}\vert}\left({\bf m}_i\times
   {\bf B}_{i}^{\rm eff}\right)\right].
\end{align} 
The LLG equation describes the dynamics of magnetization as a torque equation: the first term on the right-hand side corresponds to the precessional torque, while the second term represents the damping torque responsible for energy dissipation. In the above equation $\mathbf{m}_i$ denotes the magnetization vector of sublattice $i$. The parameters $\gamma_i$ and $\alpha_i$ represent the gyromagnetic ratio and Gilbert damping constant, respectively. The effective magnetic field $\mathbf{B}_i^{\rm eff}$ includes contributions from exchange interactions, magnetic anisotropy, demagnetizing fields and external magnetic fields. The effective field can be computed through 
\begin{align}
    \mathbf{{B}}_{i}^{\rm eff} = -\frac{\delta \Phi}{\delta \mathbf {m}_{i}},
\end{align}
where $\Phi(\mathbf{m}_{\rm Fe} , \mathbf{m}_{\rm RE})$ is the free energy of the system, given by the following expression~\cite{Mikuni2024,parchenko2016laser}:
\begin{align}
\label{Eq3}
   & \Phi(\mathbf{m}_{\rm Fe} , \mathbf{m}_{\rm RE}) = \nonumber\\
   & - \lambda \mathbf{m}_{\rm Fe} \cdot \mathbf{m}_{\rm RE} - \mathit K_{\rm Fe} \frac{\left(\mathbf{m}_{\rm Fe}\cdot \mathbf{n}\right)^{2}}{|\mathbf{m}_{\rm Fe}|^2} - \mathit K_{\rm RE} \frac{\left(\mathbf{m}_{\rm RE}\cdot \mathbf{n}\right)^{2}}{|\mathbf{m}_{\rm RE}|^2} \nonumber\\
    &-  [\mathbf{B}_{\rm THz} (t) + \mathbf{B}_{\rm ext}] \cdot(\mathbf{m}_{\rm Fe}+\mathbf{m}_{\rm RE}) \nonumber\\
    &+ \frac{\mu_0}{2}\left(\mathbf{m}_{\rm Fe}\cdot \mathbf{n} + \mathbf{m}_{\rm RE}\cdot \mathbf{n}\right)^{2}. 
\end{align}

In our model, the magnetic system consists of iron (Fe) and rare-earth (RE) sublattices, with the detailed material structure specified in a later section. The magnetization vectors of the Fe and RE sublattices are denoted by  $\mathbf{m}_{\rm Fe}$ and $\mathbf{m}_{\rm RE}$, respectively. The first term on the right-hand side of the energy expression represents the \textit{interatomic exchange interaction}, characterized by the exchange constant $\lambda$. The second and third terms correspond to the \textit{magnetic anisotropy energy} contributions from the Fe and RE sublattices, governed by the anisotropy constants $K_{\rm Fe}$ and  $K_{\rm RE}$, respectively. The fourth term accounts for the \textit{Zeeman energy} due to both a static external magnetic field $\mathbf{B}_{\rm ext}$ and a circularly polarized terahertz magnetic pulse $\mathbf{B}_{\rm THz}$. The last term captures the \textit{demagnetization energy}. In both the anisotropy and demagnetization terms, $\mathbf{n}$ denotes a unit vector along the easy axis, which we assume to be aligned with the $z$-axis throughout our calculations.

The influence of the FDT term can be accounted for through the modified effective field~\cite{Mondal2019PRB,Blank2021THz,Jingwen2022,Dutta2025jpcm}
\begin{equation}
\textbf{B}_{i}^{\rm eff} \rightarrow\left(\textbf{B}_{i}^{\rm eff}-\frac{\alpha_i a^3_i}{\gamma_i\mu_{\rm B}\mu_{\rm 0}}\frac{d\textbf{B}_{\rm THz}}{dt}\right).
    \label{Eq4}
\end{equation}
Here, $a_i^3$ denotes the unit cell volume per spin for sublattice $i$, $\mu_{\rm B}$ is the Bohr magneton, and $\mu_0$ is the permeability of the free space. This correction to the effective field was proposed theoretically in Ref.~\cite{Mondal2016}, where it was shown that the field-derivative term depends on a combination of universal constants and material-specific parameters, such as the Gilbert damping constant $\alpha$, the gyromagnetic ratio $\gamma$, and the unit cell volume per spin $a_i^3$. In contrast to conventional ferromagnets and antiferromagnets, ferrimagnets exhibit a variation in the unit cell volume per spin across different sublattices, making them particularly interesting systems to investigate the FDT-induced effects~\cite{Dutta2024}. Note that a recent experimental realization of such FDT terms has been reported in Ref.~\cite{Dutta2025}, demonstrating clear evidence of the field-derivative correction in ferrimagnetic materials.

\begin{figure}[t!]
    \centering
\includegraphics[width=1\linewidth]{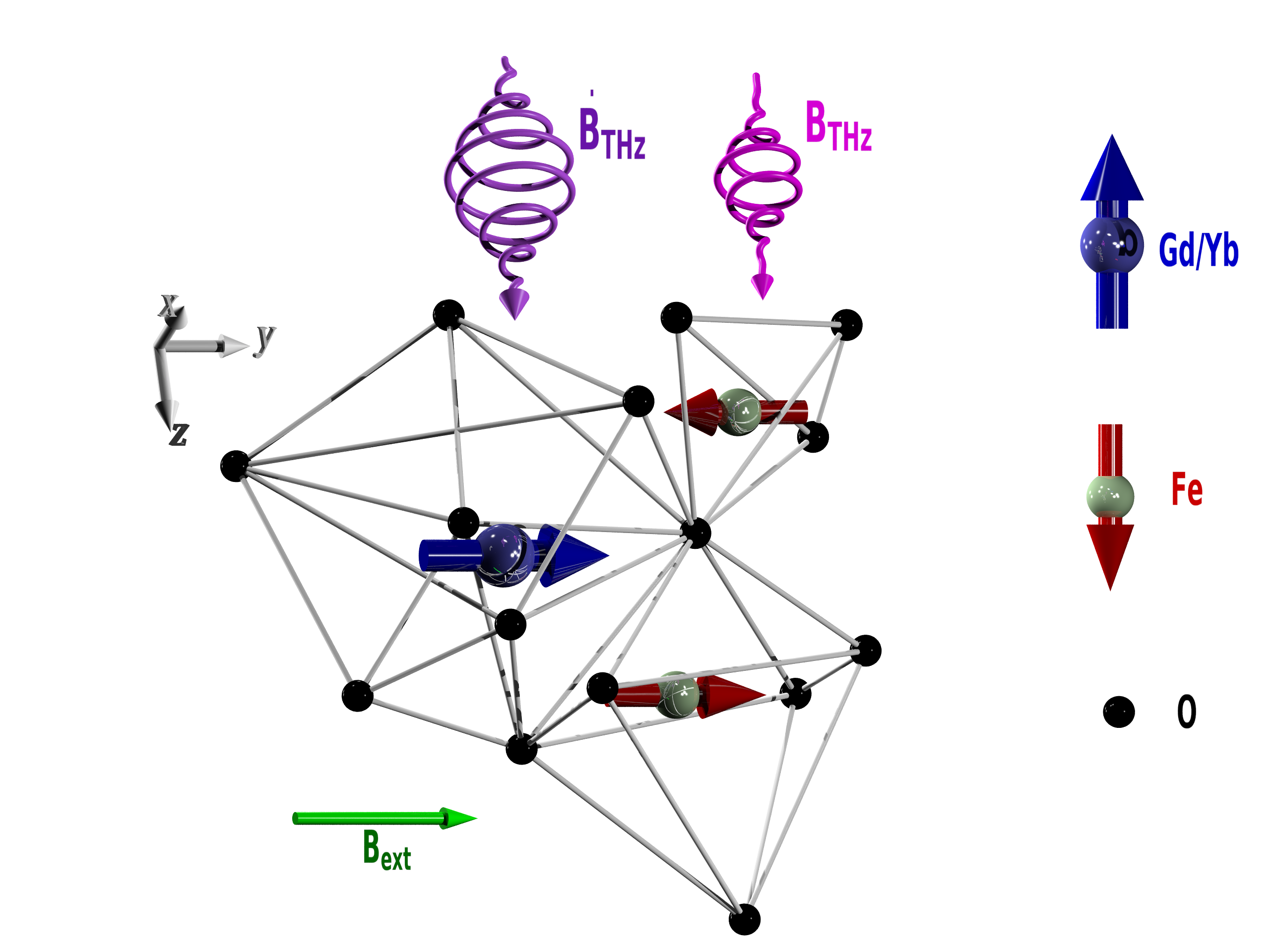}
    \caption{Schematic diagram for circularly polarized THz excitation of spins in $\rm Gd_{\frac{3}{2}}Yb_{\frac{1}{2}}BiFe_{5}O_{12}$. The rare-earth (Gd and Yb) spins are denoted in blue while the Fe spins are denoted in red. A small static magnetic field ${\bf B}_{\rm ext}$ is applied along the equilibrium direction.}
 \label{fig:1}
\end{figure}

\subsection{Details of THz pulse}
To investigate magnetization reversal, we initially employed a linearly polarized THz pulse. However, we observed no significant improvement in the required field amplitude, regardless of the presence of the FDT. We provide details on the switching diagram triggered by a linear polarized THz pulse in the Appendix~\ref{AppendixA}. To further explore the switching dynamics, we then utilized a circularly polarized magnetic pulse propagating along the $z$-direction. This was achieved by superimposing two linearly polarized magnetic pulses along the $x$- and $y$-directions, respectively, with equal amplitudes and a $\pi/2$ phase difference between them. The functional form of the employed magnetic pulses is:
\begin{align}
\label{Eq5}
    \mathbf {B}_{\rm THz}^{x}(t) = \rm B_{0} \, \cos(2\pi f_{0}\tau [\mathrm {e}^{t/\tau}-1]) \; \mathrm {e}^{-t^{2}/\sigma^{2}}\; \hat{\mathbf{x}}\,,\\
      \mathbf {B}_{\rm THz}^{y}(t) = \rm B_{0} \, \sin(2\pi f_{0}\tau [\mathrm {e}^{t/\tau}-1]) \; \mathrm {e}^{-t^{2}/\sigma^{2}}\; \hat{\mathbf{y}}\,.
      \label{Eq6}
\end{align}
The THz pulse computed from these equations resemble the pulses used in earlier experiment~\cite{Kampfrath2011,Hirori2011,Dutta2025}. Here $\rm B_{0}$ is the amplitude of the THz field, and $\rm f_{0}$ is the peak frequency of the THz pulse, the value of which is taken to be equal to the Kaplan-Kittel resonance frequency of the material. The chirp time $\tau$ represents the instantaneous frequency change over time. We have used a positive chirp in our simulations. The width of the THz pulse is related to $\sigma$ via the relation full width at half maxima (FWHM) = 0.833$\sigma$. In our simulations, Eqs.~(\ref{Eq5}) and~(\ref{Eq6}) characterize a circularly polarized terahertz pulse, with both the chirp time and the pulse width lying within the picosecond range i.e., $\tau = 2.6$ ps and $\sigma = 1$ ps. In the latter part of the article, we vary the THz pulse width for computing the magnetization switching diagrams. We note that while the above-mentioned THz fields exert ZT on the spins, the time derivatives of such fields, i.e., $\dot{\mathbf {B}}_{\rm THz}^{x}(t)$ and $\dot{\mathbf {B}}_{\rm THz}^{y}(t)$ exert FDT on the magnetic spins.     
\begin{figure*}[tbh!]
    \centering
    \includegraphics[width=0.497\textwidth]{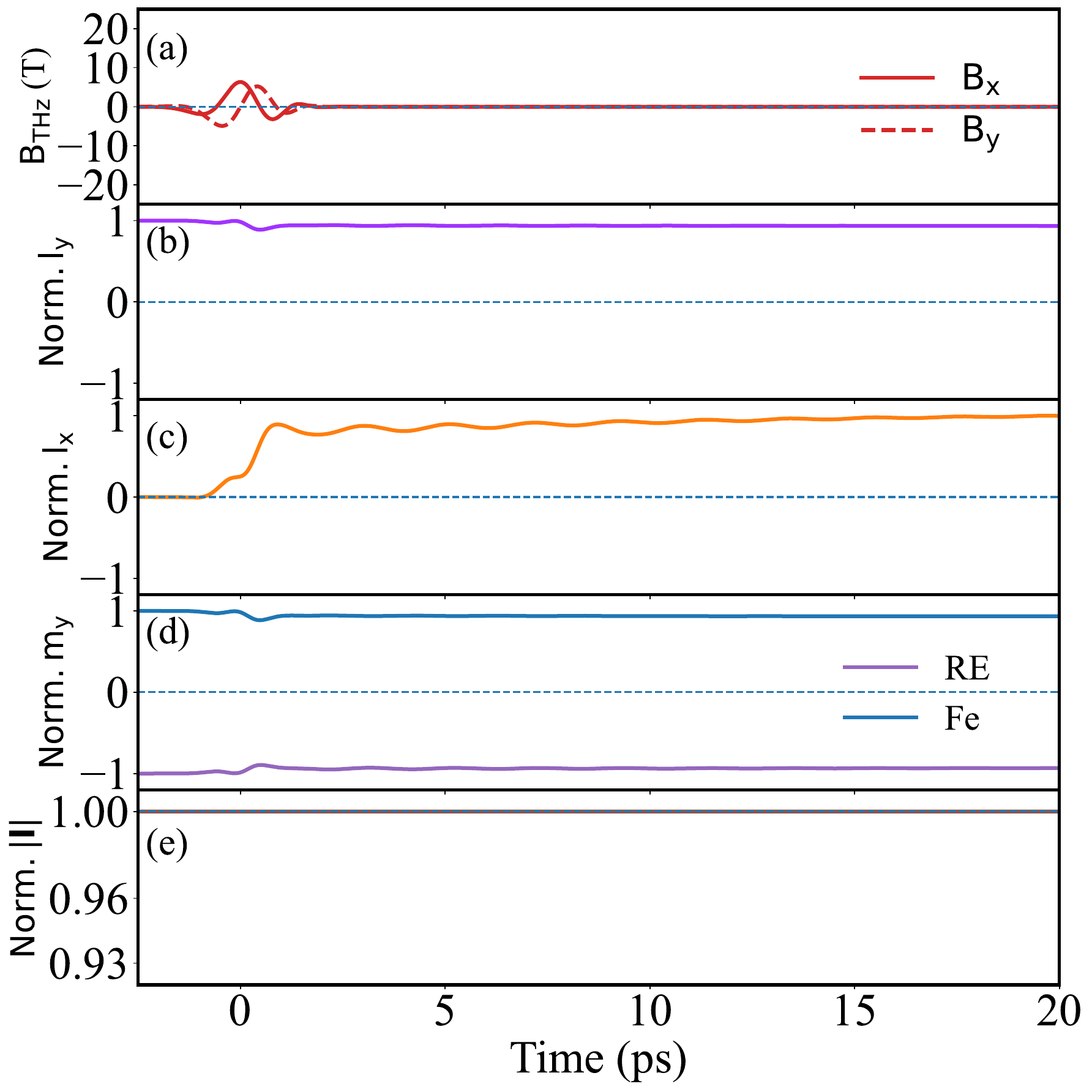}
    \includegraphics[width=0.497\textwidth]{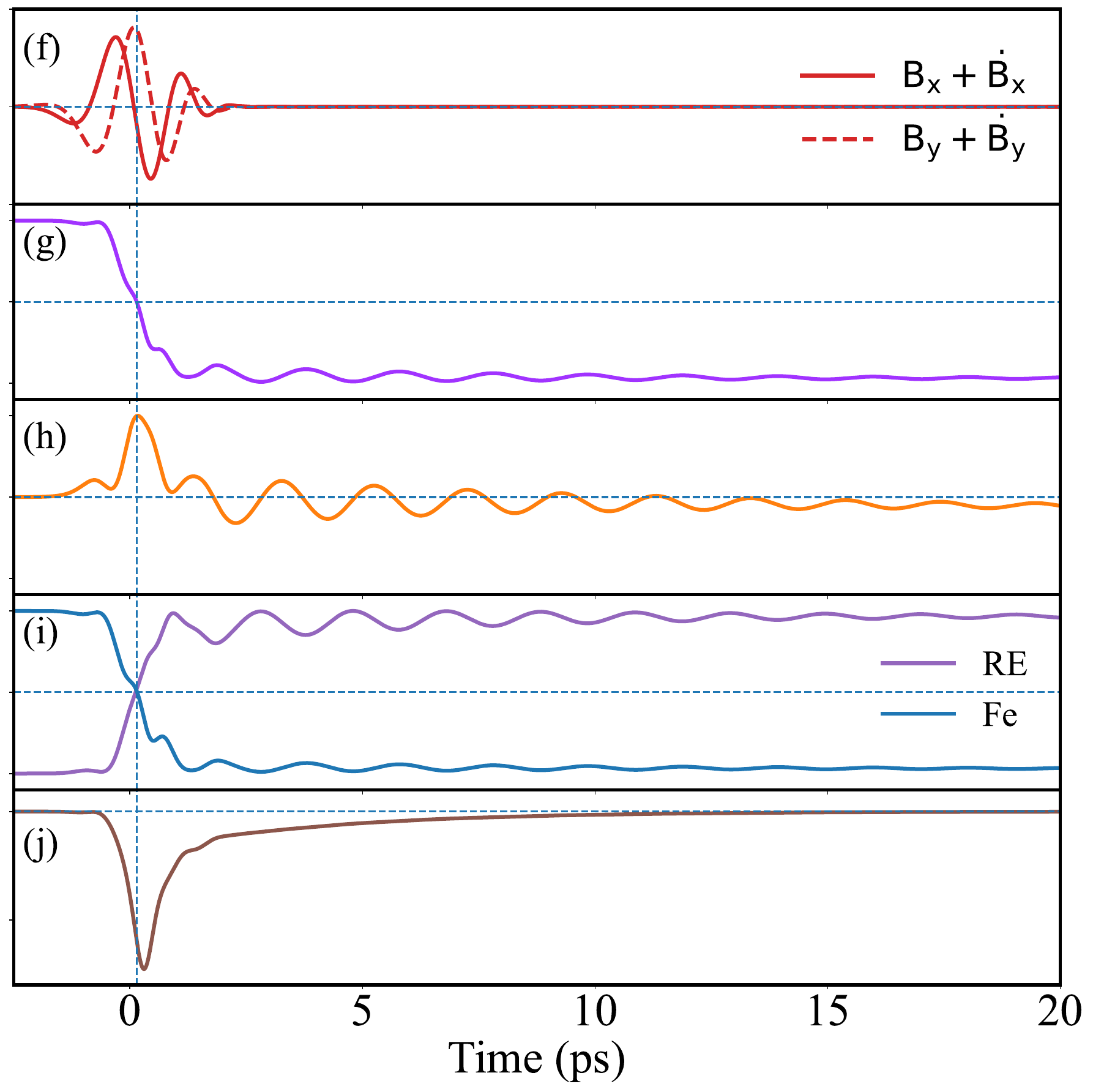}
    \captionof{figure}{Simulation results {of the magnetization dynamics without FDT (Left Panel) and with FDT (Right Panel) for incident THz magnetic fields of ${\rm B_0 }= 6.35$\,T, $\sigma = 1$\,ps and $\tau = 2.6$\,ps.} (a) and (f) show { the $x$- and $y$-components of the circularly-polarized THz pulse}, (b) and (g) show the $y$-component of the N\'eel vector, {which gets switched in presence of  FDT, and not switched in the absence of FDT}. (c) and (h) show the $x$-component of the N\'eel vector while (d) and (i) show magnetization dynamics of the individual sublattices. The normalized $\vert {\bf l}\vert$ in the absence and presence of FDT are shown in (e) and (j), respectively.} 
    \label{fig:2}
\end{figure*}

\subsection{Material specifications}
The system studied is the ferrimagnetic compound $\rm Gd_{\frac{3}{2}}Yb_{\frac{1}{2}}BiFe_{5}O_{12}$. A schematic of material's unit cell along with the incident circularly-polarized THz fields is illustrated in Fig.~\ref{fig:1}. The ferrimagnetic material crystallizes in the cubic garnet structure. The garnet unit cell consists of three distinct magnetic sublattices: one formed by the RE ions and two by the Fe ions occupying tetrahedral and octahedral sites. The Fe sublattices exhibit strong antiferromagnetic coupling between themselves, while the resulting net Fe moment couples antiparallel to the RE moment located at the dodecahedral sites. This interplay leads to a ferrimagnetic ground state that persists at room temperature \cite{Satoh2012,parchenko2013wide}. In our schematic diagram in Fig.~\ref{fig:1}, the RE (Gd$^{3+}$, Yb$^{3+}$) moments are represented by blue arrows while the Fe$^{3+}$ are shown by red arrows. The nonmagnetic oxygen atoms are denoted by the black dots.   
The equilibrium magnetization of both the RE and Fe atoms is aligned along the $y$-direction. To mimic the experimental situation, an external magnetic field of $\rm {B_{ext}} = 120$\,mT, also oriented along the same direction, is applied (shown by the green arrow in Fig.~\ref{fig:1}). The RE and Fe sublattices have equilibrium magnetization of $-50 \times 10^{3}$\,J/Tm$^{3}$ and $140\times 10^{3}$\,J/Tm$^{3}$~\cite{parchenko2016laser}, respectively, making the system ferrimagnetic. The material exhibits a Kaplan–Kittel resonance frequency of 0.48\,THz and a Gilbert damping constant of 0.02~\cite{parchenko2013wide,Satoh2012}, making it well-suited for observing the effects of the FDT in our case. Since the magnetization arises from two distinct atomic species in the ferrimagnetic $\rm Gd_{\frac{3}{2}}Yb_{\frac{1}{2}}BiFe_{5}O_{12}$, the unit cell volume per spin $a^3_i$ becomes a significant parameter. For the RE sublattice, this volume is $a^3_{\rm RE} = 8.5 \times 10^{-29}$\,m$^3$, while for the Fe sublattice, it is $a^3_{\rm Fe} = 1.2 \times 10^{-28}$\,m$^3$~\cite{Blank2021THz,Dutta2024,Dutta2025}. The exchange constant of the material is $\lambda = -1930 \times 10^{-7}$\,T$^2$m$^3$/J, and the anisotropy constants for both sublattices are 1000\,J/m$^3$~\cite{Dutta2024,Dutta2025}. Further, we employed fundamental physical constants, including the Bohr magneton $\rm \mu_{\rm B} = 9.274 \times 10^{-24}$\,J/T, gyromagnetic ratio of $1.76 \times 10^{11}$\,s$^{-1}$T$^{-1}$, and vacuum permeability $\rm \mu_{\rm 0} = 4\pi \times 10^{-7}$\,Vs/Am.

\section{Results and Discussions}
Our simulations were performed at $T = 0$ K, considering two magnetic sublattices, namely Fe and RE. Consequently, thermal effects were not included in the simulations. Instead of using optical or laser pulses, we employ a single-cycle circularly polarized THz pulse with a frequency of $\rm f_0 = 0.48$ THz following Eqs.~(\ref{Eq5}) and~(\ref{Eq6}).  We define the N\'eel vector as $\mathbf{ l} = {\bf m}_{\rm Fe}-  {\bf m}_{\rm RE}$, where $\mathbf{l} = \rm l_x \mathbf{\hat{x}} + \rm l_y \mathbf{\hat{y}} + \rm l_z \mathbf{\hat{z}} $.
The primary objective of our investigation is to understand how the FDT terms facilitate magnetization switching on a picosecond timescale and to elucidate the underlying mechanisms. To this end, we have analyzed the time evolution of $\rm l_x$ to examine how the effective torque influences magnetization dynamics along the $x$-direction. Additionally, we have examined the normalized components of $\rm l_y$ and $\rm m_y$ of the individual sublattices to clearly demonstrate the magnetization reversal process.
Further, we also show the dynamical variation of the normalized N\'eel vector to explain the underlying switching mechanism.
We use a circularly polarized THz pulse having field amplitude ${\rm B_0} = 6.35$\,T. The comparison of the THz pulse fields signifies that the FDT enhances the effective THz field amplitude [see Fig.~\ref{fig:2}(f)] compared to without the FDT in Fig.~\ref{fig:2}(a). Such results are consistent with earlier investigations as well~\cite{Mondal2019PRB,Dutta2024,Dutta2025}.

The equilibrium magnetization is oriented along the $y$ axis. As a result, the applied THz magnetic field component ${\rm B}^y_{\rm THz}(t)$ initially exerts no torque on the magnetization. In contrast, the ${\rm B}^x_{\rm THz}(t)$ component generates a torque along the $z$-direction, causing the magnetization to tilt away from its equilibrium position. Once the magnetization is deflected, ${\rm B}^y_{\rm THz}(t)$ begins to contribute to the total torque, further influencing the magnetization dynamics. Due to the weak torque without FDT, the Néel vector undergoes magnetization dynamics without a complete switching and relaxes back to its equilibrium orientation once the THz pulse is turned off [see Fig.~\ref{fig:2}(b)]. In contrast, when the FDT is included, the equilibrium magnetization undergoes a complete and deterministic reversal to the opposite direction, as shown in Fig.~\ref{fig:2}(g). This switching is stable and does not relax back to the original equilibrium state after the THz pulse is removed. 

To illustrate the torque exerted on the magnetization, we plot the $x$-component of the N\'eel vector in Fig.~\ref{fig:2}(c) and~\ref{fig:2}(h). In the absence of the FDT, the torque reaches its maximum at longer timescales, after the THz pulse has been removed. In contrast, when the FDT is included, the torque reaches its peak much earlier, around 0.15 ps. This moment is indicated by a vertical dotted line, emphasizing that magnetization switching occurs within this timescale. We observe that the norm. ${\rm l_x}$ also oscillates around the equilibrium, however, at longer timescales ($\sim$ 500\,ps), owing to the GHz ferromagnetic resonance (FMR) mode. We note that such FMR modes can be seen in all the components of the N\'eel vector, irrespective of the inclusion of the FDT.    

\begin{figure}[t!]
    \centering
    \includegraphics[width=1\linewidth]{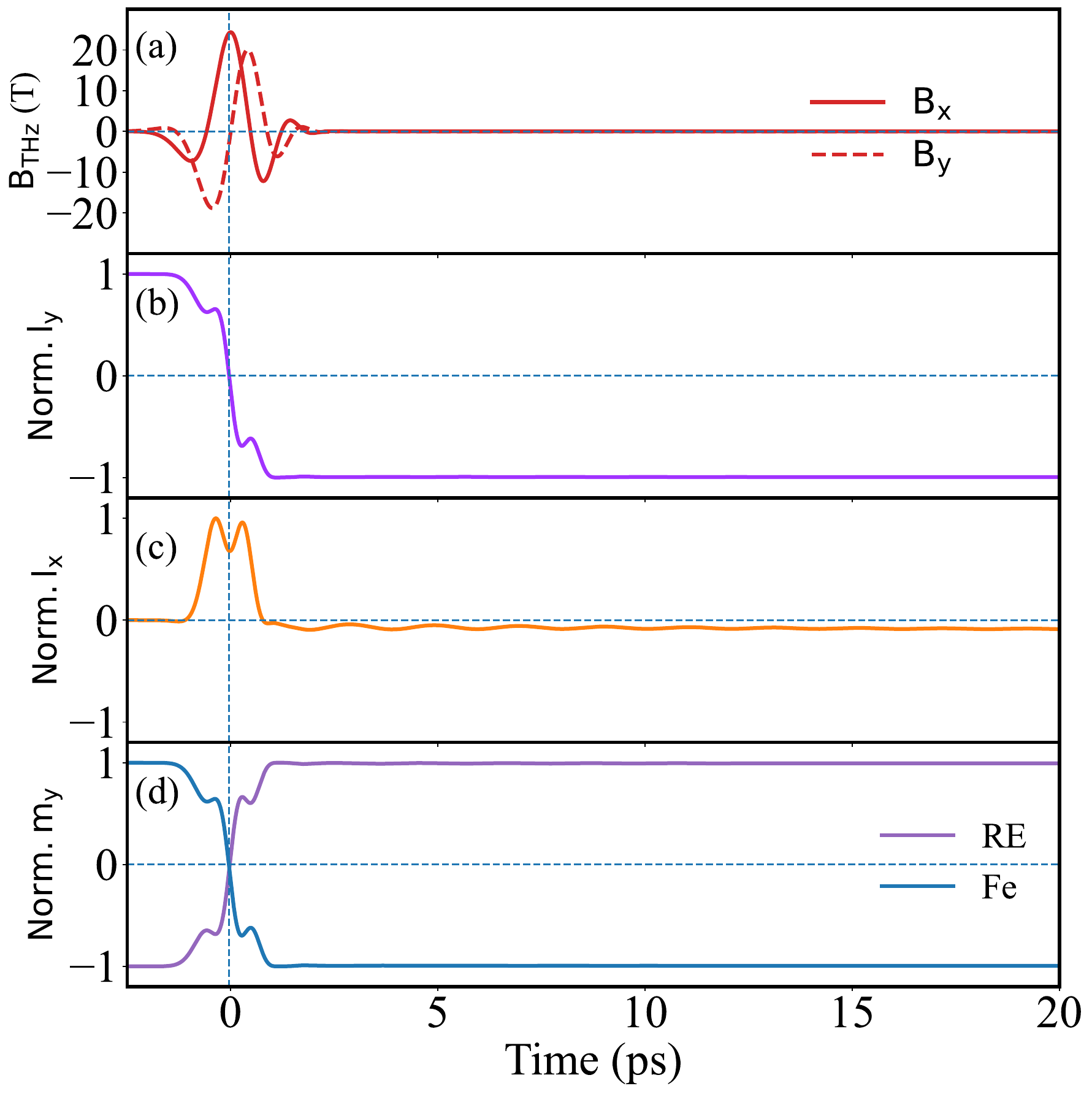}
    \caption{Magnetization reversal without the FDT at a field amplitude of B$_0 = 24.35$\,T, $\sigma = 1$\,ps and $\tau = 2.6$\,ps. (a) The circularly polarized field components, (b) the N\'eel vector in the $y$-direction, (c) the N\'eel vector in the $x$-direction, (d) individual sublattice magnetization dynamics in the $y$-direction ("RE" and "Fe" sublattices) are shown.}
    \label{fig:3}
\end{figure}

From these observations, we conclude that in the presence of the FDT, the torque saturates at the peak of the magnonic oscillations in the $x$-direction. The resulting excess torque is transferred to the equilibrium magnetization direction, initiating magnetization reversal in both sublattices. These switching of the individual sublattices are been shown in Fig.~\ref{fig:2}(d) and~\ref{fig:2}(i). Note that such magnetization reversal is rather different from the laser-induced ferrimagnetic switching in GdFeCo, where a transient ferromagnetic-like state mediates the switching~\cite{Radu2011,Wienholdt2013,MorenoPRB2017,Gerlach2017}. In our case, the transient ferromagnetic-like state is not observed, meaning there is no transfer of angular momentum between the sublattices involved in the switching process. 

To further investigate the magnetization reversal, we have calculated the dynamical change of the normalized magnitude of the N\'eel vector. Without the FDT, the normalized magnitude of the N\'eel vector does not deviate from its maximal equilibrium value [see Fig.~\ref{fig:2}(e)]. In contrast, with the FDT, the N\'eel vector shows a dip during the presence of the circularly polarized THz pulse. Once the pulse is switched off, the N\'eel vector relaxes back to its equilibrium. Without the FDT, the two sublattices respond symmetrically and experience identical torques. In contrast, the presence of the FDT breaks this symmetry due to the different magnetizations and unit cell volume per spin, causing the sublattices to experience unequal torques. This results in a phase and amplitude mismatch between the magnonic oscillations of the two sublattices, most prominently in the $x$- and $z$-directions. Such scenario leads to a reduction in the instantaneous magnitude of the normalized N\'eel vector and eventual magnetization reversal with the FDT. A similar magnetization dip can also be observed in the dynamical change of the angle between two sublattices ($\Theta$), as well as the exchange energy between the two sublattices. A more detailed discussion has been provided in Appendix~\ref{appendixc}.

Next, we investigate whether the magnetization reversal can occur in the absence of the FDT. Note that THz magnetization reversal in antiferromagnetic and ferrimagnets has been reported without the FDT in Ref.~\cite{Sonke2012}. The magnetization switching in NiO was found to require a very high magnetic field, typically greater than 18 T, irrespective of the THz pulse width~\cite{Sonke2012}. In our investigation of ferrimagnetic $\rm Gd_{\frac{3}{2}}Yb_{\frac{1}{2}}BiFe_{5}O_{12}$, we find the magnetization switching occurs at field amplitude B$_0$ = 24.35\,T with $\sigma = 1$ \,ps. Such switching results are been shown in Fig.~\ref{fig:3}. Similar to the magnetization reversal with the FDT, we observe the maximum of torque during the presence of the THz pulse as shown in Fig.~\ref{fig:3}(c), which then initiates the magnetization reversal. However, we note that the phase of the {normalized} ${\rm l_x}$ involved in Figs.~\ref{fig:3}(c) and~\ref{fig:2}(h) are different. This difference is due to the presence of the FDT in the latter case. The individual sublattice switching has been shown in Fig.~\ref{fig:3}(d). To summarize, we find that the presence of the FDT significantly reduces the required THz magnetic field strength for inducing magnetization switching.  

\begin{figure*}[t!]
    \centering
    \includegraphics[width=0.497\textwidth]{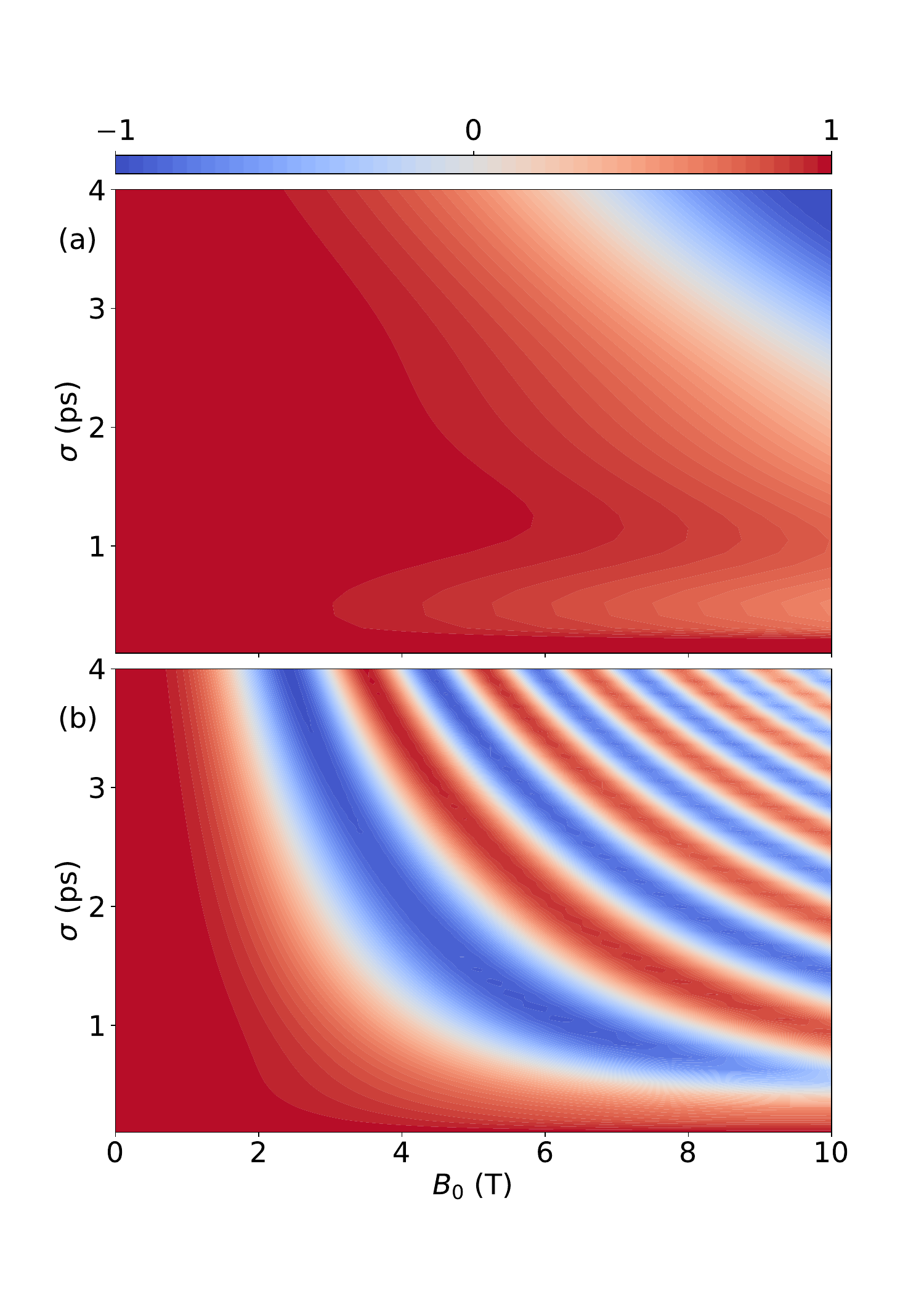}
    \includegraphics[width=0.497\textwidth]{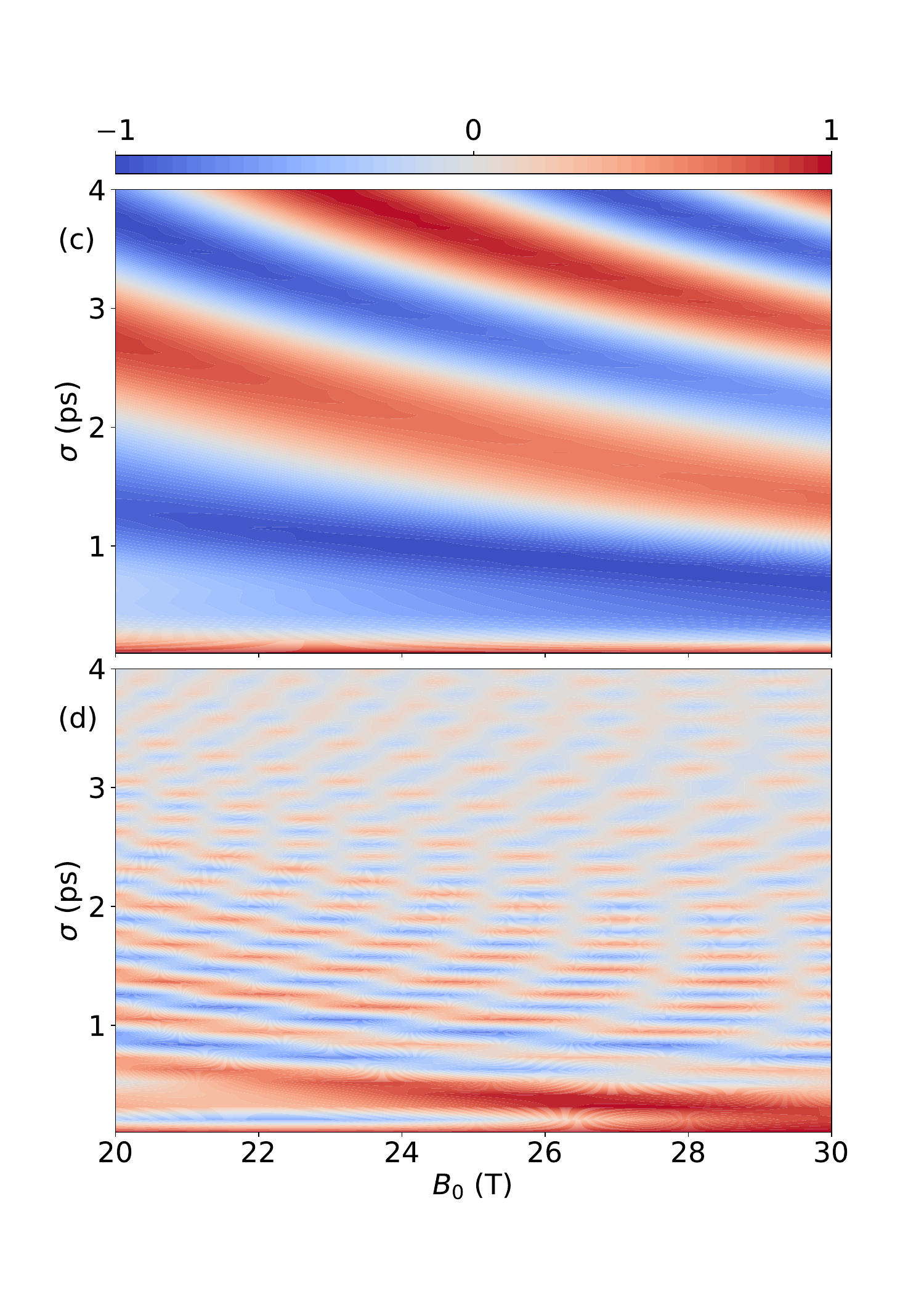}
    \vspace{-15pt}
    \caption{The value of normalized ${\rm l_y}$ after circular polarized THz excitation of $\rm Gd_{\frac{3}{2}}Yb_{\frac{1}{2}}BiFe_{5}O_{12}$. While the red areas signify norm. ${\rm l_y} = 1$, meaning the magnetization is not reversed, the blue areas denote norm. ${\rm l_y} = -1$ specifying the magnetization has reversed. The regions with other colors signify demagnetization. (a) Without FDT and (b) with FDT show the switching and non-switching areas within the field strength within B$_0 = 0 - 10$\,T with pulse width range up to $\sigma = 4$\,ps. (c) Without FDT and (d) with FDT show the same areas at a higher THz field strength within B$_0 = 20 - 30$\,T.}     
    \label{fig:4}
\end{figure*}

Although the magnetization reversal happens at a lower THz magnetic field with the FDT, such reversal also depends on the {temporal width} of the THz pulse. To investigate further, we compute the value of norm. ${\rm l_y}$ after the THz pulse excitation of $\rm Gd_{\frac{3}{2}}Yb_{\frac{1}{2}}BiFe_{5}O_{12}$ with picosecond circular pulses having different field strengths and pulse durations. These results are summarized in Fig.~\ref{fig:4}. In the red area, the N\'eel vector shows the magnetization dynamics upon excitation of the THz pulse; however, the system relaxes back to its equilibrium state represented by norm. ${\rm l_y} = 1$. In contrast, the blue areas represent the N\'eel vector have been switched i.e., the value of normalized ${\rm l_y} = -1$ after the THz pulse excitation. The Fig.~\ref{fig:4}(a) and~\ref{fig:4}(c) denote the switching and non-switching areas without the FDT, while~\ref{fig:4}(b) and~\ref{fig:4}(d) shows the same with the FDT. The field strength B$_0$ is varied from 0\,T to 10\,T in Figs.~\ref{fig:4}(a) and~\ref{fig:4}(b), and from 20\,T to 30\,T in Figs.~\ref{fig:4}(c) and~\ref{fig:4}(d). The pulse width of the circularly polarized THz pulse is varied up to $\sigma$ = 4\,ps in these plots.

We observe that magnetization reversal occurs at higher THz magnetic fields and longer pulse widths, which is consistent with previous studies~\cite{Sonke2012}. This behavior is evident in the upper-right corner of Fig.~\ref{fig:4}(a). However, in the presence of the FDT, switching occurs at lower THz field strengths and shorter pulse durations. For instance, when the pulse width is set to $\sigma = 1$\,ps, magnetization reversal is observed at field strengths above 6\,T, as also discussed in Fig.~\ref{fig:2} (Right Panel). Moreover, increasing the pulse width further reduces the required THz field for switching. As an example, a clear switching regime is observed at ${\rm B_0} = 2.94$\,T with a pulse width of $\sigma = 4$\,ps when the FDT is included. Nonetheless, it is important to note that there exists a lower bound on the THz magnetic field strength necessary to induce switching, even for extremely long pulse durations. Overall, the inclusion of the FDT leads to a significantly broader switching region compared to the case without the FDT. We note that there exist regions where the final state is demagnetized, i.e., it does not return to its equilibrium magnetization state ($\mathrm{norm.}\ {\rm l_y} = 1$) nor transition to the reversed magnetization state ($\mathrm{norm.}\ {\rm l_y} = -1$). These demagnetized regions serve as boundaries between the switching and non-switching regimes.

Next, we analyze the switching and non-switching regimes at higher field strengths, specifically in the range of 20\,T to 30\,T. These regimes are presented in Figs.~\ref{fig:4}(c) and~\ref{fig:4}(d), corresponding to the cases without and with FDT, respectively. At higher values of $B_0$, a broader switching regime is observed even in the absence of FDT. However, the extent of demagnetized regions also increases significantly at such high field strengths, indicating that overly strong THz pulses can lead to breakdown of the magnetic order rather than controlled reversal. In the presence of FDT, which effectively enhances the THz field strength, Fig.~\ref{fig:4}(d) shows an almost complete suppression of distinct switching and non-switching regions. Instead, due to the elevated effective field, only demagnetized regions are observed. This suggests that while the FDT facilitates switching at lower field strengths, it may push the system into an unstable regime at higher fields, where the energy input is sufficient to destroy the sublattice magnetization entirely. Therefore, precise control of both the pulse strength and pulse width in the presence of FDT is essential for achieving deterministic magnetization switching.

\section{Conclusion}
We computationally demonstrate that magnetization reversal in ferrimagnets can be achieved using a circularly polarized THz pulse, aided by the field derivative torque, at a lower THz field amplitude of $6.35$\,T with a pulse width of $\sigma = 1$\,ps. This required field amplitude is significantly lower than the $24.35$\,T observed without including FDT, having the same pulse width. We contemplate the switching mechanism of magnetization by plotting different components of the N\'eel vector and sublattice magnetization. Due to the FDT, the two sublattices in a ferrimagnet experience unequal torque leading to the magnetization switching. Moreover, our results show that for $\rm\sigma = 4$\,ps, magnetization reversal occurs at a field amplitude of $2.94$\,T when the FDT term is included. This represents a significant improvement over previously reported field amplitudes for ferrimagnetic switching, where such switching was reported with THz field amplitudes more than 18\,T~\cite{Wienholdt2013}.

Due to the high spin-orbit coupling, the ferrimagnetic material $\rm Gd_{\frac{3}{2}}Yb_{\frac{1}{2}}BiFe_{5}O_{12}$ exhibits a relatively large Gilbert damping parameter of $\alpha = 0.02$. Moreover, the Kaplan-Kittel exchange mode in this material lies at 0.48\,THz. These two key features are essential for the manifestation of the FDT, observed in this system. Although our simulations incorporating the FDT demonstrate a substantial reduction in the THz field strength required to achieve magnetization switching, the generation of high-field, circularly polarized THz pulses remains experimentally challenging~\cite{YuanPRL,Song2020,Bogatskaya2020}. {Several schemes have been reported for the generation of strong field circular THz pulses taking the combined advantage of tilted pulse front THz generation using LiNbO$_3$ and THz metasurfaces~\cite{Li_2024}, from wavelength-scale plasma oscillator~\cite{Wu2008} or by the application of Mach-Zehnder interferometry~\cite{Ueda2025}. MV/cm scale generation of linearly polarized THz pulses have also been reported~\cite{Sell2008,Junginger2010,Hoffmann2011,Mondal2017ol} that can be used to convert to circular polarization using one of the schemes mentioned above.} Building upon this work, circularly polarized THz pulses in conjunction with ferrimagnetic materials possessing optimized combinations of Gilbert damping, gyromagnetic ratio, and unit cell volume per spin may be identified. Such material-pulse pairings could pave the way for the development of faster and more energy-efficient memory devices in future spintronic technologies.

\section{Acknowledgment}
R.M. acknowledges SERB-SRG via Project No. SRG/2023/000612 and the faculty research scheme at IIT (ISM) Dhanbad, India, under Project No. FRS(196)/2023-2024/PHYSICS. A.D. and S.P. acknowledge the support from DAE through the project Basic Research in Physical and Multidisciplinary Sciences via RIN4001. S.P. also acknowledges the start-up support from DAE through NISER and SERB through SERB-SRG via Project No. SRG/2022/000290.  The authors acknowledge V. Chakraborty and D. N. Basu for the fruitful discussions.
\appendix
\section{Effect of linearly polarized THz pulse}
\label{AppendixA}
\begin{figure}[t!]
    \centering
    \includegraphics[width=1.2\linewidth]{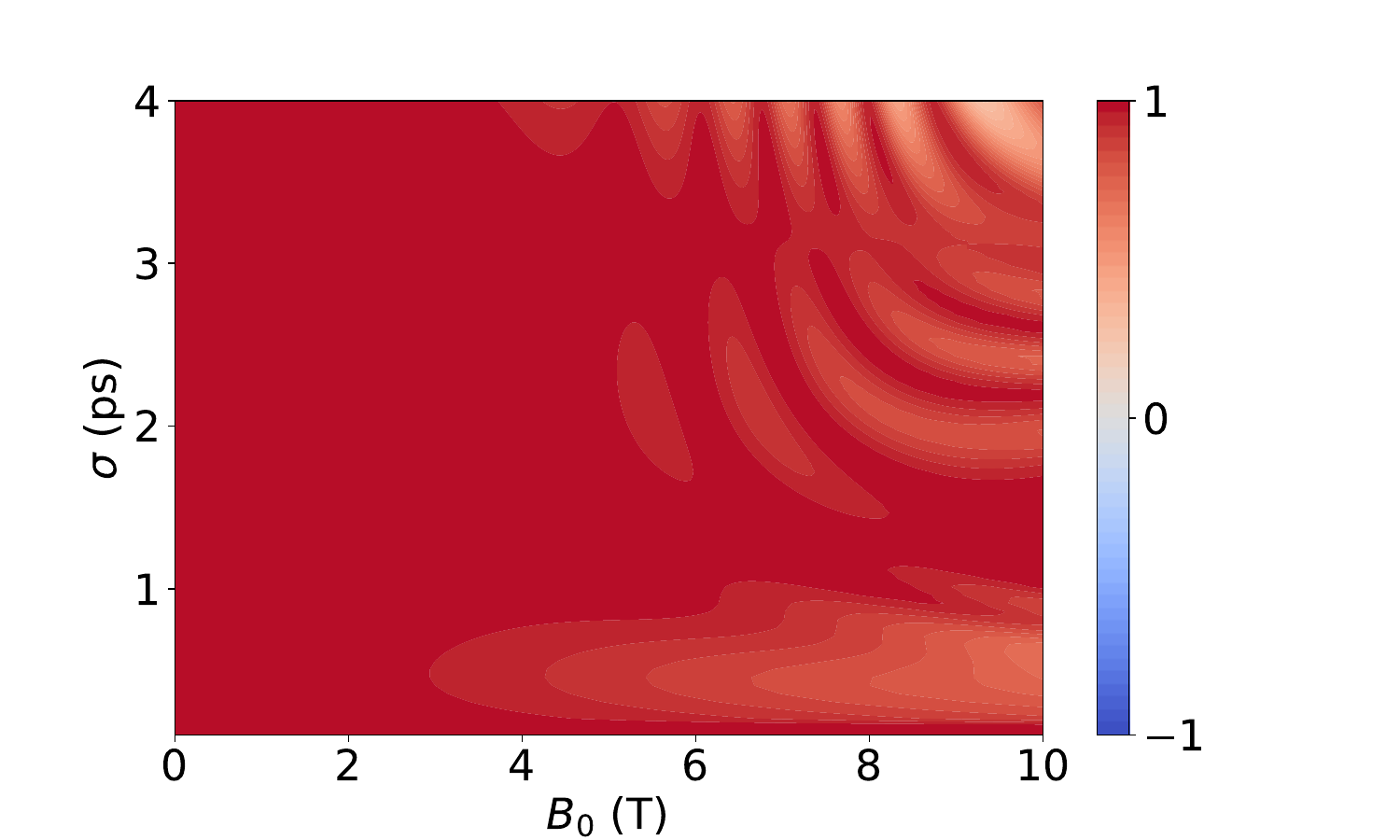}
    \caption{The switching diagram with linearly polarized THz pulse in the presence of FDT. The red areas signify norm. ${\rm l_y} = 1$, meaning the magnetization is not reversed.}
        \label{fig:5}
\end{figure}
In a previous work regarding the experimental observation of FDT~\cite{Dutta2024,Dutta2025}, linear polarized pulses were used in the experiments and corresponding theoretical simulations. We discuss the FDT-induced switching in case of the linear polarized THz pulse. In particular, we consider the THz pulse propagating along the $z$-direction having an expression 
\begin{align}
        \mathbf {B}_{\rm THz}(t) = \rm B_{0} \, \cos(2 \pi f_{0}\tau [\mathrm {e}^{t/\tau}-1]) \; \mathrm {e}^{-t^{2}/\sigma^{2}} \hat{x}
\end{align}
Fig.~\ref{fig:5} shows the switching of N\'eel vector with the FDT after the excitation of linear polarized THz pulse. We observe that the magnetization switching does not occur within the field strength of 10\,T and pulse width of 4\,ps. Thus, we conclude that the FDT does not improve the magnetization reversal in the case of excitation with the linear polarized THz pulse.  

For the lower values of $\rm\sigma$ and ${\rm B_0}$, the plot appeared to be completely red, which means that the final value of normalized $\rm{l_y}$ does not deviate from its initial value. As one moves away from the origin of the plot, an irregular distribution of unevenly closed shapes in mixed shades of red and white becomes apparent. This indicates that for higher values of $\rm\sigma$ and ${\rm B_0}$, the normalized value of $\rm{l_y}$ deviates from its equilibrium value. However, it does not correspond to the magnetization switching. 

We therefore continue to explore the magnetization reversal following excitation by a circularly polarized THz pulse. A comparison between Fig.~\ref{fig:4}(a) and Fig.~\ref{fig:4}(b) reveals that the application of FDT significantly enhances the magnetization reversal in this ferrimagnetic material. This improvement highlights the crucial role of material-dependent parameters (Gilbert damping, gyromagnetic ratio, unit cell volume per spin) in facilitating ultrafast spin dynamics.  

\section{Observed dip during THz pulse excitation}
\label{appendixc}
We have seen that the magnetization reversal is associated with a dip in the norm. $\vert {\bf l}\vert$ during the THz pulse excitation, in the presence of the FDT. Such a dip can be seen in Fig~\ref{fig:2}(j). In this appendix, we provide a detailed analysis regarding the dip of the N\'eel vector. 
\begin{figure}[t!]
\centering
\includegraphics[width=0.42\textwidth]{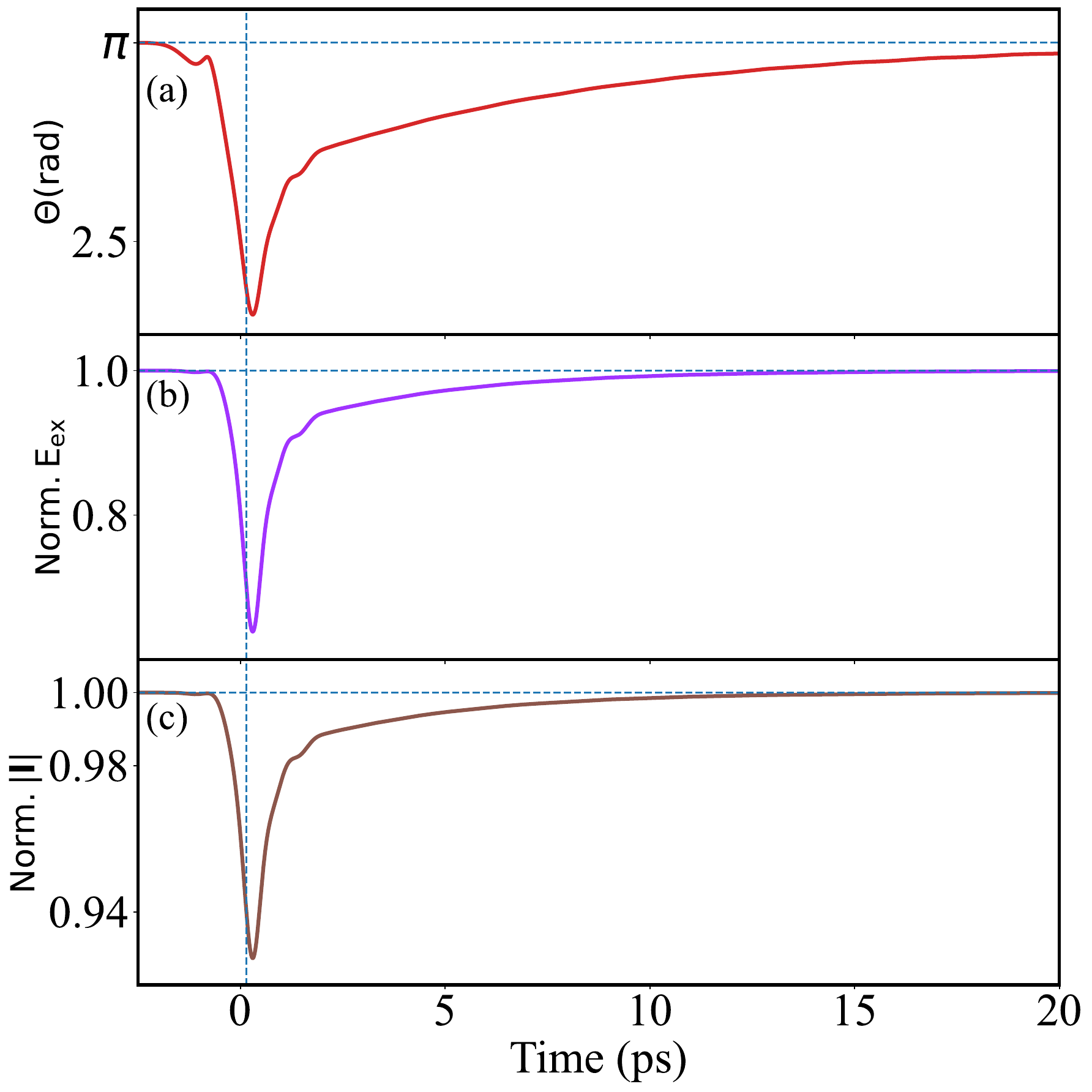}
    \caption{The dynamics of (a) $\Theta$, the instantaneous angle between two sublattices, (b) normalized exchange energy (${\rm E_{ex}}$), and (c) norm. magnitude of the N\'eel vector in the presence of the FDT-induced magnetization switching with circularly polarized THz pulse having ${\rm B_0} = 6.35$\,T and $\sigma = 1$\,ps.} 
    \label{fig:6}
\end{figure}
We compute the N\'eel vector via 
\begin{align}
    \mathbf{l} = \mathbf{m}_{\rm Fe} - \mathbf{m}_{\rm RE} \Rightarrow 
    \mathbf{\lvert l \rvert}^{\rm 2}  =  {\lvert \mathbf{m}_{\rm Fe} - \mathbf{m}_{\rm RE} \rvert}^{\rm 2}
\end{align}
which results in
\begin{align}
    \mathbf{\lvert l \rvert}^{\rm 2}  & =  {\lvert \mathbf{m}_{\rm Fe} \rvert}^{\rm 2} + {\lvert \mathbf{m}_{\rm RE} \rvert}^{\rm 2} - 2 \,\, { \mathbf{m}_{\rm Fe} } \cdot { \mathbf{m}_{\rm RE}}\nonumber\\
    & = {\lvert \mathbf{m}_{\rm Fe} \rvert}^{\rm 2} + {\lvert \mathbf{m}_{\rm RE} \rvert}^{\rm 2} - 2 \,\, { \vert \mathbf{m}_{\rm Fe} \vert }  { \vert \mathbf{m}_{\rm RE}\vert }\cos \Theta
\end{align}
where $\Theta$ is the angle between the two sublattices. On the other hand, $ { \mathbf{m}_{\rm Fe} } \cdot { \mathbf{m}_{\rm RE}}$ also signifies the considered exchange energy. Thus, we have computed the time evolution of $\Theta$, the normalized exchange energy (norm. E$_{\rm ex}$), and the normalized magnitude of $\vert {\bf l} \vert$ to demonstrate that they all exhibit consistent dynamical behavior. The results have been shown in Fig.~\ref{fig:6}. 

The dotted vertical line in Fig.~\ref{fig:6} highlights the time when the magnetization switching occurs. 
The Figs.~\ref{fig:6}(a),~\ref{fig:6}(b) and~\ref{fig:6}(c) show $\Theta$, norm. ${\rm E_{ex}}$ and norm. $\vert {\bf  l}\vert$ as a function of time in the presence of FDT with a THz field amplitude of 6.35\,T and pulse width $\sigma = 1$\,ps. All these plots show a significant dip while the THz pulse is in action and returns to the norm. ${\vert {\bf l}\vert} = 1$ after the THz pulse is switched off. Note that Fig.~\ref{fig:6}(c) is the same as Fig.~\ref{fig:2}(j) shown earlier. During the switching process with the FDT, the angle between the two sublattices does not always remain 180$^{\circ}$. Similarly, the exchange energy also changes as well which facilitates the magnetization switching. Such dips do not significantly occur without the FDT, and hence the magnetization switching is not achieved without FDT at lower THz field strength.



\providecommand{\noopsort}[1]{}\providecommand{\singleletter}[1]{#1}%

\end{document}